\documentclass[12pt]{iopart}

\usepackage{graphicx}
\usepackage{dcolumn}
\usepackage{bm}
\usepackage{hyperref}
\usepackage{subfigure}
\usepackage{amssymb}
\usepackage{color}
\usepackage{mathrsfs}
\usepackage{times}
\usepackage{subeqnarray}
\usepackage{cases}
\usepackage{epstopdf}
\usepackage{marvosym}

\begin{document}

\title[Pinning control of social fairness in the Ultimatum game]{Pinning control of social fairness in the Ultimatum game}

\author{Guozhong Zheng$^1$, Jiqiang Zhang$^2$, Zhenwei Ding$^2$, Lin Ma$^1$ and Li Chen$^1$$^,$$^*$}

\address{$^1$ School of Physics and Information Technology, Shaanxi Normal University, Xi'an 710062, P. R. China}
\address{$^2$ School of Physics and Electronic-Electrical Engineering, Ningxia University, Yinchuan 750021, P. R. China}
\address{$^*$Author to whom any correspondence should be addressed.}
\ead{chenl@snnu.edu.cn}
\vspace{10pt}
\begin{indented}
\item[]31 December 2022
\end{indented}

\begin{abstract}
Decent social fairness is highly desired both for socio-economic activities and individuals, as it is one of the cornerstones of our social welfare and sustainability. How to effectively promote the level of fairness thus becomes a significant issue to be addressed. Here, by adopting a pinning control procedure, we find that when a very small fraction of individuals are pinned to be fair players in the Ultimatum Game, the whole population unexpectedly evolves into the full fairness level. The basic observations are quite robust in homogeneous networks, but the converging time as a function of the pinning number shows different laws for different underlying topologies. For heterogeneous networks, this leverage effect is even more pronounced that one hub node is sufficient for the aim, and a periodic on-off control procedure can be applied to further save the control cost. Intermittent failures are seen when the pinning control is marginally strong, our statistical analysis indicates some sort of criticality. Our work suggests that the pinning control procedure could potentially be a good strategy to promote the social fairness for some real scenarios when necessary.

\end{abstract}


%
%
\maketitle
%
%

\section{Introduction}
In his bestseller, \emph{Capital in the Twenty-first Century}~\cite{Piketty2014Capital}, Thomas Piketty reported that half of the wealth is owned by the top one percent of the population globally, and the bottom 70\% only owns 3\% of the wealth. The concern of inequality or social unfairness has been so far attracting many attentions from economists, political scientists, psychologists, and even physicists. The key questions to be addressed are: what are the mechanisms for the fairness emergence, and how to improve the level of fairness for those unsatisfactory scenarios.

Most previous studies of fairness are based upon the Ultimatum game (UG) model~\cite{Guth1982An}. Two persons are to divide a sum of money, and they act as the role of proposer and responder alternately. The proposer first proposes an allocation plan, and the responder chooses either to agree or reject. If the agreement is reached, the money will be divided according to the plan; otherwise both get nothing.

Within the paradigm of \emph{Homo economicus}~\cite{Simon1957Models,Samuelson2005Economics}, people are interest maximizers, and a low fairness outcome is expected in the UG. Since ``something is better than nothing", the responder will accept any small amount of money; with this knowledge, the proposer should propose the smallest possible offer to the responder to keep the largest amount to his/her own interest.  However, a lot of behavior experiments have revealed that we humans have a remarkable preference for fairness. When playing the UG, most offers are between 40\% and 50\% of the total amount, very few offers are below 20\% and will be rejected by responders with more than half a chance~\cite{Guth1982An, Thaler1988Anomalies,Bolton1995Anonymity, Roth1995The, Guth2014More}. Evidences in non-human species also support this observation~\cite{Brosnan2003Monkeys,Proctor2013Chimpanzees}.

Different mechanisms are later proposed to understand the emergence of fairness. One theoretical finding is that once the population is structured, the fairness level is better than the well-mixed case~\cite{Page2000The,Kuperman2008The,Sinatra2009The}. Besides, the factors of fame and reputation~\cite{Nowak2000Fairness, Andre2011Social}, empathy~\cite{Sanchez2005Altruism, Page2002Empathy}, and noise~\cite{Binmore1994An, Gale1995Learning} are also found to have potentially important impacts on the evolution of fairness.
While quite a few studies aim to decipher the mechanism behind fairness emergence, much less effort has been seen to improve it when the level of fairness is unsatisfactory. Undoubtedly, there are a lot of examples of unfair divisions where the fairness promotion is highly desired, such as the global vaccine coverage on the COVID-19 pandemic, where more than one million lives might have been saved if the vaccines had been shared more equitably~\cite{moore2022retrospectively}.

\begin{figure*}[htbp]
\centering
\includegraphics[width=0.32\textwidth]{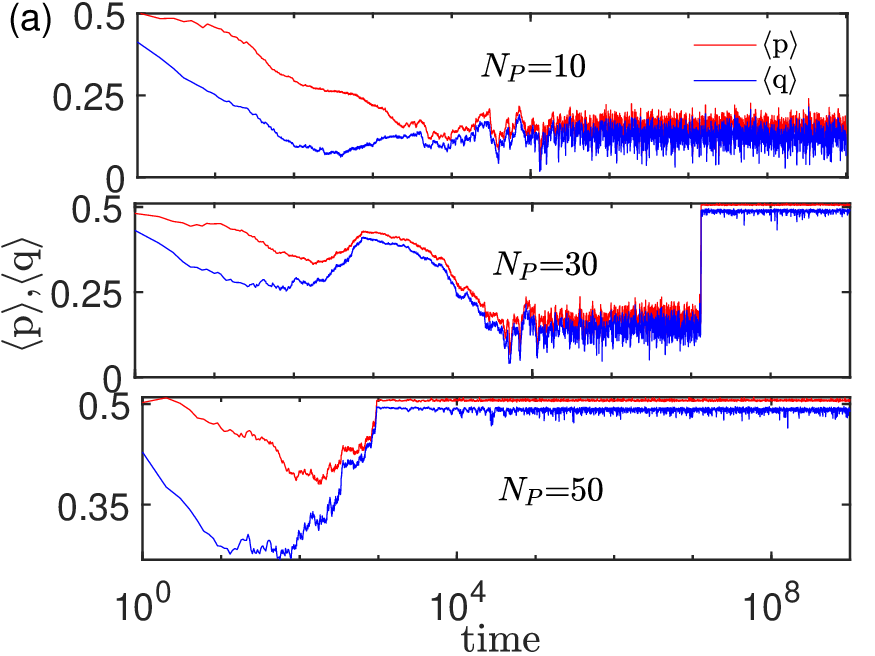}
\includegraphics[width=0.32\textwidth]{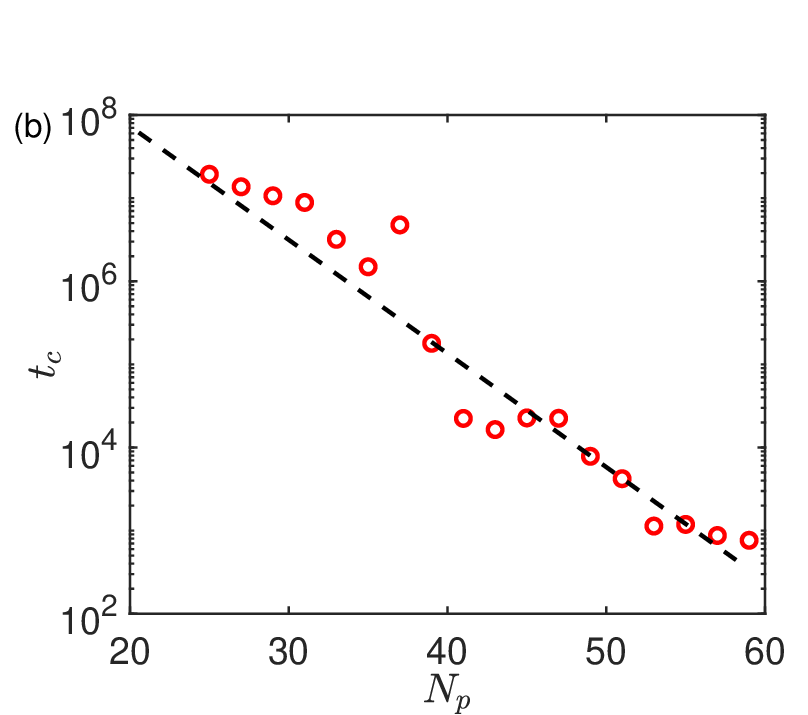}
\includegraphics[width=0.32\textwidth]{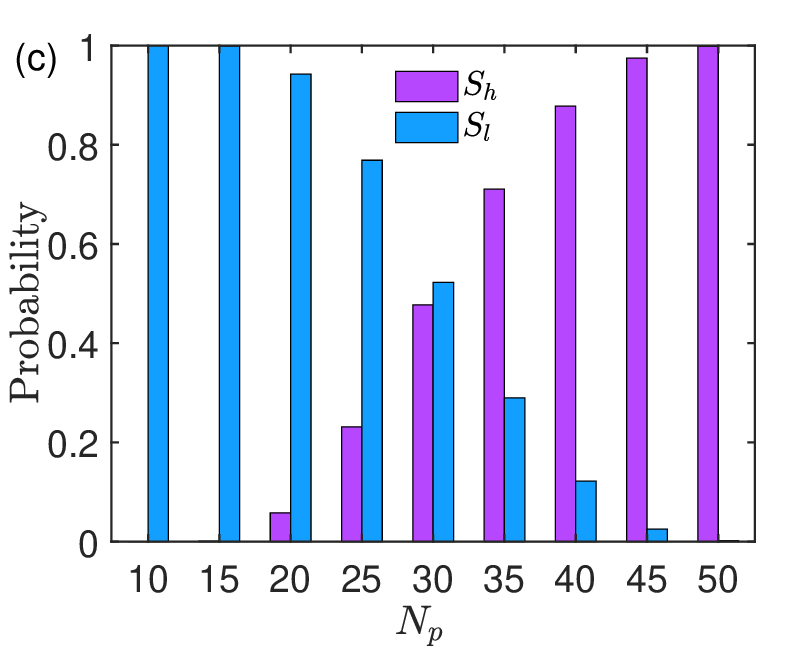}
\caption{(Color online) The evolution of fairness on the $1d$ lattice under pinning control.
(a) Typical time series for the average offer $\langle p\rangle$ and the acceptance threshold $\langle q\rangle$ for three different pinning number $N_p=10,30,50$.
(b) The converging time $t_{c}$ towards the high fairness state versus  $N_p$, 30 ensemble averages are conducted for each $N_p$. The simulation time for each run is $10^8$ steps that three orders longer than the convergence time in the classic case, we only average those realizations that successfully evolve into the high fairness state in the end.
(c) The probabilities for the population evolving into the high and low fairness state $S_{h,l}$ versus $N_p$, $10^4$ ensemble averages are conducted for each $N_p$.
Parameters: $N=1000$, $K=40$.
}
\label{fig:lattice}
\end{figure*}

In Ref.~\cite{zheng2022probabilistic}, when people probabilistically choose to be ``good Samaritans", the fairness level typically exhibits a first-order phase transition and there is a nucleation process behind that explains the emergence of the high fairness level. Observations are also made in the evolution of cooperation where the overall prevalence can also be driven by a small number of zealots who choose unconditional cooperation~\cite{Masuda2012Evolution,James1988Free}. Inspired by these work, here we try to promote the level of fairness from the control point of view, specifically, via the pinning control procedure. Originally, this idea is derived from the pinning effect in condensed matter physics~\cite{Di2005Pinning}, but later it is widely applied to a variety of contexts, including the turbulence~\cite{Tang2006Controlling}, plasma instability, reaction-diffusion systems~\cite{Wang2016Pinng}, and synchronization on complex networks~\cite{Sorrentino2007Controllability,Yu2013Synchronization,YU2009On,Zhou2008Pinning}, as well as the problem of resource allocation optimization~\cite{Zhang2013Controlling,Zhang2016Controlling}, etc. The general idea behind this is that by pinning only a small number of individuals, the impact will disproportionately drive the whole system to the target states.

Specifically, we apply the method of pinning control to structured populations playing the UG model, by pinning a small number of individuals to be fair players, we succeed in achieving a high level of fairness for the population as a whole. While the leverage effect is similar, the converging time shows different dependence on the pinning number. More pronounced leverage effect is revealed in heterogeneously structured populations due to the structural heterogeneity.
The specific questions to be addressed are: How many nodes need to be pinned? How long it takes to reach the fairness state? For those heterogeneous networks, which nodes should be selected for the optimal control effect? 

The paper is organized as follows: we introduce our model in Sec. \ref{sec:model}. In Sec. \ref{sec:results} we show results for the population with 1d lattice and Erd\H{o}s-R\'enyi random networks. In Sec. \ref{sec:networks}, Barab\'asi-Albert scale-free networks are also examined to study the impact of the structural heterogeneity. Finally, we conclude our work together with some discussions in Sec. \ref{sec:discussion}.

\section{Model}
\label{sec:model}
We consider a structured population of $N$ individuals playing the Ultimatum game, the underlying topology will be specified later on. As a rule of thumb~\cite{Guth1982An,Debove2016Models}, the sum of the amounts of money to be divided is set to be the unity. The strategy for player $i$ is given by $S_i\!=\!(p_i,q_i)$, where $p_i$ is the \emph{offer} when acting as a proposer and $q_i$ is the \emph{acceptance threshold} when acting as a responder. Obviously, $p_i,q_i\in[0,1]$, usually their initial values are chosen randomly in that range.

Now, when two players $i,j$ encounter each other and they are assumed in the role of proposer and responder with equal chance,  the expected payoff for player $i$ versus $j$ is as follows

\begin{equation}
  \pi_{ij}= \left\{
   \begin{array}{lr}
	1-p_i+p_j,\quad\quad \mbox{if}\ p_i\geqslant q_j\ \mbox{and}\  p_j\geqslant q_i,\\
	1-p_i,\quad\quad\quad\quad\, \mbox{if}\ p_i\geqslant q_j\ \mbox{and}\  p_j< q_i,\\
	p_j,\quad\quad\quad\quad\quad\;\;\:  \mbox{if}\  p_i<q_j\:\mbox{and}\  p_j\geqslant q_i,\\
	0,\quad\quad\quad\quad\quad\;\;\:\:\,    \mbox{if}\  p_i<q_j\ \mbox{and}\  p_j<q_i.\\
  \end{array}
   \right.
   \label{eq:rule}
\end{equation}

At the very beginning, $N_p$ individuals are randomly selected as the pinning sites, they will stick to the fair strategy $S_h\!\equiv\!(p_h,q_h)$ with $p_h\!=\!q_h\!=\!C_h=0.5$ throughout the evolution. This ``irrational" behavior may happen when they are supported financially by external parties like the government or for other incentives~\cite{Camerer2003Advances,Loewenstein2003The,Sanfey2003The,Rilling2011The}. The rest $N-N_p$ players evolve according to the following synchronous updating rule.

These players are initially assigned with a random strategy $p_i,q_i\in[0,1]$ uniformly and independently. Next, each player interacts with all their nearest-neighbors, obtains their payoffs according to the Eq.~({\ref{eq:rule}}), and sums them up $\pi_i=\sum_{j\in\Omega_i}\pi_{ij}$, where $\Omega_i$ denotes the neighborhood of player $i$. Finally, the strategy reproduction is proportional to the payoffs in the neighborhood, each player updates its strategy according to the Moran rule~\cite{Roca2009Evolutionary}

\begin{equation}
	w(S_j\rightarrow S_i)=\frac{\pi_j}{\sum_{j\in\Omega'_i}\pi_j}, \label{eq:moran}
\end{equation}
where $w(S_j\rightarrow S_i)$ is the probability that player $i$ imitates the strategy of player $j$ in its neighborhood, including itself ($\Omega'_i=\Omega_i \cup \{i\}$). To the end, each imitation is subject to a small noise $p_i=p_j+\delta p$ and $q_i=q_j+\delta q$, where $(\delta p,\delta q)\in[-\varepsilon,\varepsilon]$, and we set $\varepsilon=0.001$.

To measure the population fairness level, we calculate the average offer $\langle p\rangle$ and the average acceptance level $\langle q\rangle$ of the population as our two order parameters. When $(\langle p\rangle,\langle q\rangle)\approx(0.5,0.5)$, i.e. $S_h$, the population reaches the ideal half-half split, which is always desired. It's worth mentioning that the value of $q_i$ is almost always slightly smaller than $p_i$ in the long run (see Appendix for a detailed analysis), thus tracking one of them is usually sufficient for monitoring the evolution.

Obviously, when $N_p=0$, the case is then reduced to the classic model~\cite{Kuperman2008The} that has been extensively studied previously; when $N_p\rightarrow N$, a high fairness level is trivially expected.  We are primarily interested in the following questions: is it possible to promote the overall fairness by pinning very few  individuals? And if yes, what fraction is needed?

\section{Results}
\label{sec:results}
\subsection{$1d$ Lattices}

\begin{figure}[htbp]
\centering
\includegraphics[width=0.45\textwidth]{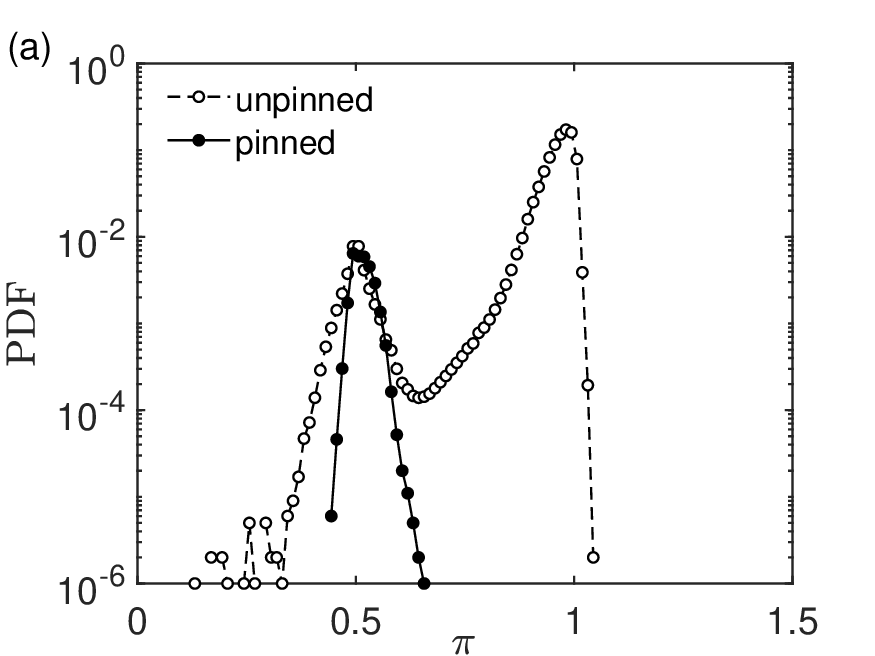}
\includegraphics[width=0.45\textwidth]{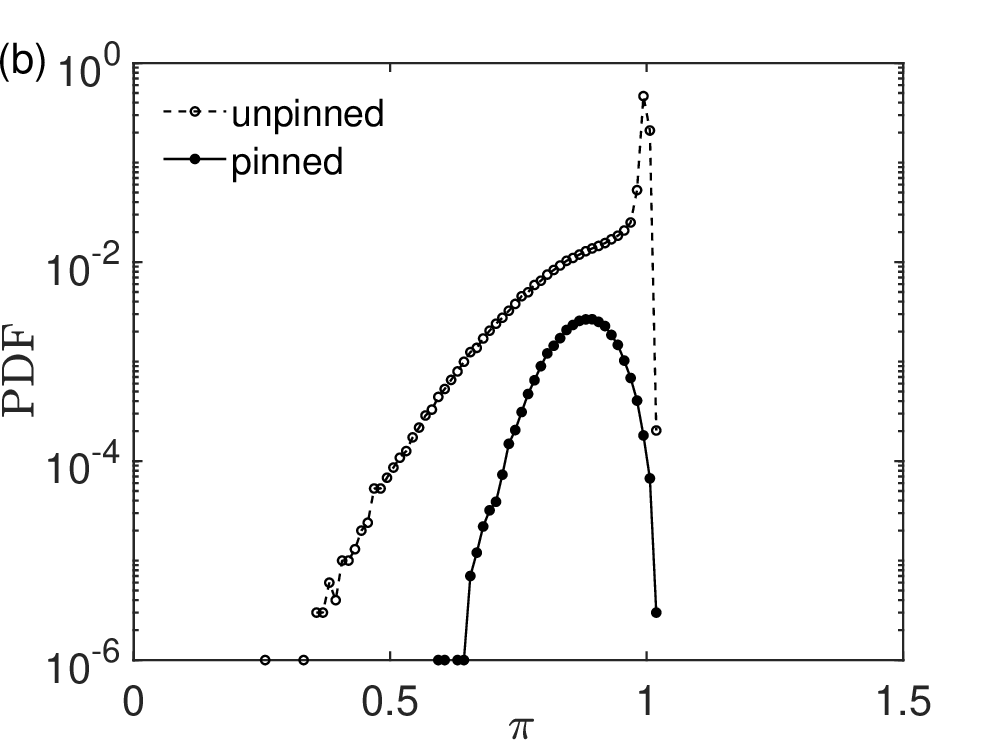}
\caption{The probability density distribution of payoff on 1d lattice within the bistable region for two different realizations. (a) and (b) are for scenarios where the high fairness state is unsuccessfully and successfully reached, respectively. The solid/open points correspond to the pinned/unpinned individuals.
Parameters: $N_p\!=\!30$, $N=1000$, and the degree $K=40$.
}
\label{fig:Earnings}
\end{figure}

We first report the numerical results of a one-dimensional lattice population with $N=1000$, each individual is connected to $K/2$ neighbors on either side, where the degree $K=40$ in our study.
Fig.~\ref{fig:lattice}(a) shows that when the number of pinning individuals is relatively small $N_{p}=10$, the fairness level is almost the same as the classic case (i.e., $N_p=0$)~\cite{Kuperman2008The}; individuals are willing to offer only $10\%\sim20\%$ share to their partners on average, denoted as the typical low fairness strategy $S_l$.  Interestingly, a further increase to $N_p=30$ potentially leads to a sudden jump into the half-half split state --- all with the fully fair strategy $S_h$. However, this success is probabilistic. Depending on the initial condition, some realizations evolve into the high fairness state, while others stay in the low fairness state.
Keep increasing $N_{p}$ to 50, the population evolves into the high level of fairness for all realizations, all individuals with $S_h$.
This means that, when $5\%$ of the population are pinned, this control procedure guarantees the overall fairness level to the ideal level $S_h$ for the whole population.

To see how long it takes to reach the high fairness state, the average converging time $t_c$ is computed for different $N_p$, shown in Fig.~\ref{fig:lattice}(b). We see that $t_c$ decreases as a function of $N_p$ approximated as $t_c\sim e^{-\beta N_p}$ with $\beta\approx 0.148$. The trend is understandable, since the more pinning individuals are, the greater the impact of pinning control is, thus the faster convergence to the target state.
Furthermore, the probabilities towards $S_h$ and $S_l$ are also given for several cases of $N_p$ in Fig.~\ref{fig:lattice}(c). It shows that with the increase of $N_p$, the probability towards $S_h$ is gradually increased as expected, since the basin of attraction for the high fairness state expands. At $N_p=30$, this evolution to $S_h$ is at half chance, while this probability reaches $100\%$ for the case of $N_p=50$.

From the practice point of view, the pinning control might need financial support externally, it's natural to ask what's the difference between pinned and unpinned individuals in terms of payoffs, since this is related to how much cost is needed. Thus, we provide the PDF of their payoffs respectively for both unsuccessfully and successfully pinned to the high fairness state, in Fig.~\ref{fig:Earnings}(a) and (b). When failed to approach the target state, Fig.~\ref{fig:Earnings}(a) shows that the payoff of pinned players is significantly lower than unpinned on average, additional cost might be needed to compensate for their loss. But once the fairness state is reached,  the two payoffs are very close,  no compensation is required afterwards.

As to why the pinning control works, the mechanism is largely the same as our previous work~\cite{zheng2022probabilistic}. Physically, the dynamical process is similar to a nucleation phenomenon in the phase transition of matter states~\cite{1987Introduction}, where the pinned sites act as nucleation cores. Once some fair clusters are formed and grow successfully at the initial stage, they merge with each other till the whole population are with the strategy $S_h$. But, if the initial growth fails, the clusters around these pinned sites cannot reach each other, no coalition is expected and the majority of the population is in a low fairness state with the strategy $S_l$. A mean-field theory based on the replicator equation~\cite{Roca2009Evolutionary,Taylor1978Evolutionary} is developed in Ref.~\cite{zheng2022probabilistic} that captures the physical picture and is also applicable here.

\subsection{Erd\H{o}s-R\'enyi random networks}
To see whether the above observations are dependent on the underlying structures of the population, we also examine two complex topologies: Erd\H{o}s-R\'enyi (ER) random networks~\cite{Bollobas2001random} and  Barab\'asi-Albert (BA) scale-free networks~\cite{Barabasi1999Emergence}, which represents the typical homogeneous and heterogeneous networks, respectively.

\begin{figure}[tbp]
\centering
\includegraphics[width=0.5\linewidth]{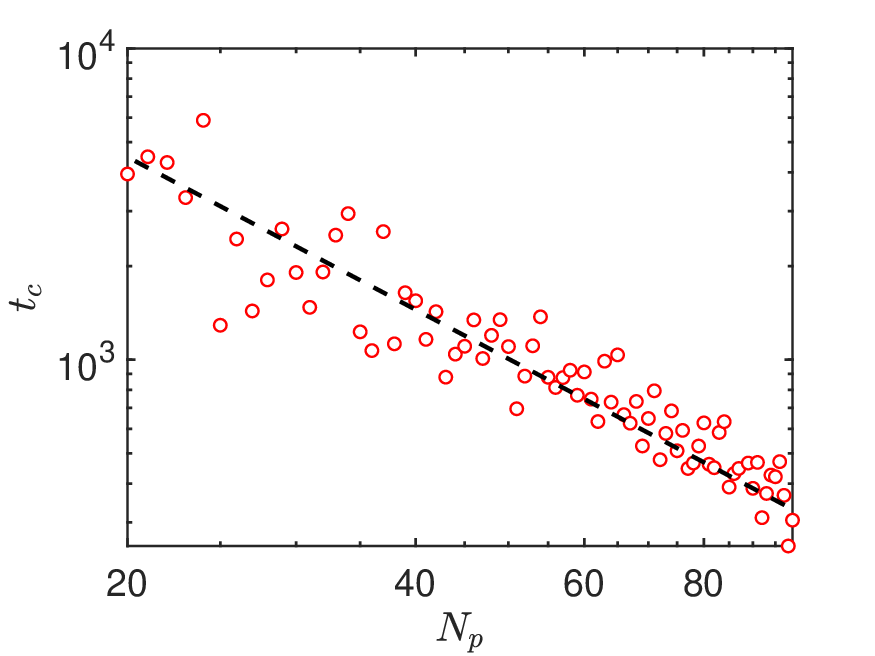}
\caption{
(Color online) The converging time $t_c$ versus the pinning number $N_p$ on ER networks, over 30 ensemble averages for each $N_p$.
Parameters: $N=1000$, the average degree $\langle k\rangle=4$.
}
\label{fig:ER}
\end{figure}

Note that the ER networks we used are with the same network size $N$ as above and the average degree $\langle k\rangle=4$, the pinning nodes are also randomly chosen and fixed.
Similar observations are made that: i) pinning too few ($N_p\lesssim 20$) nodes fails to work; ii) pinning the intermediate number ($20\lesssim N_p\lesssim80$) yields probabilistic success towards the high fairness state; iii) when $N_p\gtrsim80$ the pinning control can always drive the whole population to $S_h$.
Interestingly, Fig.~\ref{fig:ER} shows that the dependence of the converging time $t_c$ on the pinning number $N_p$ now presents approximately a power law $t_c\sim N^{-\gamma}_p$, with $\gamma\approx1.46$. This dependence is different from the results in the lattice in Fig.~\ref{fig:lattice}(b), where a very long $t_c$ is required for relatively small $N_p$ but this is not case on ER networks. The reason lies in the fact that different pinned nodes acting as nucleation cores are located far away from each other on $1d$ lattice, thus taking quite some time to get merged; while ER networks have much shorter average shortest distance, it's much faster for these cores to do so.

\begin{figure*}[tbp]
\centering
 \includegraphics[width=0.8\textwidth]{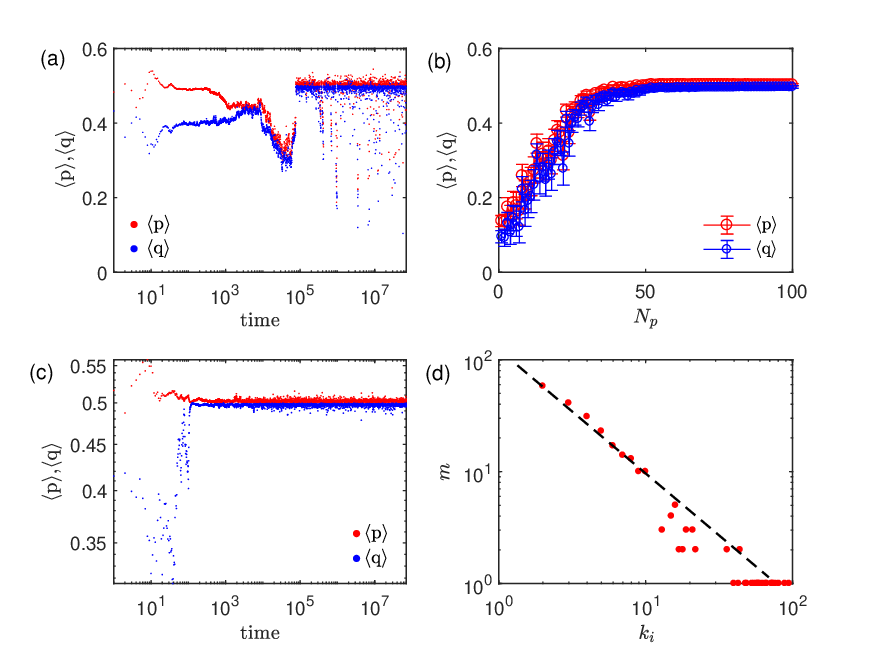}
\caption{(Color online) The evolution of fairness on BA scale-free networks under pinning control.
Typical time series of $\langle p\rangle$ and $\langle q\rangle$ by either randomly pinning $N_p$=30 nodes (a), or just the largest hub (c); we see both reach the high fairness state, but strong fluctuations are present in (a).
(b) Continuous transition of $\langle p\rangle$ and $\langle q\rangle$ are shown versus the number of randomly pinning nodes $N_p$. The error bars are standard deviations of the time series after a transient of $8\times10^6$ steps.
(d) The minimal number $m$ of pinned individuals required to achieve the fair state versus the nodes' degree $k_i$,  a power law is approximated as $m \sim k_{i}^{-1.4}$.
Parameters: $N=1000$, $\langle k\rangle=4$.
}
\label{fig:BAtime}
\end{figure*}

Put together, our study shows that the effect of pinning control approach seems robust to the underlying topology of the homogeneously structured population, the leverage effect always can be seen by pinning not too small fraction of individuals. In fact, even in the well-mixed population, qualitatively the same results are expected according to our previous mean-field treatment~\cite{zheng2022probabilistic} and this is confirmed numerically (data not shown). Once the population is structured in a heterogeneous way, the leverage effect is even more pronounced with a different picture, as shown in the following section.
\section{Scale-free networks}
\label{sec:networks}

\subsection{Pinning random individuals vs the largest hub}

\begin{figure*}[bpth]
\centering
\includegraphics[width=0.32\textwidth]{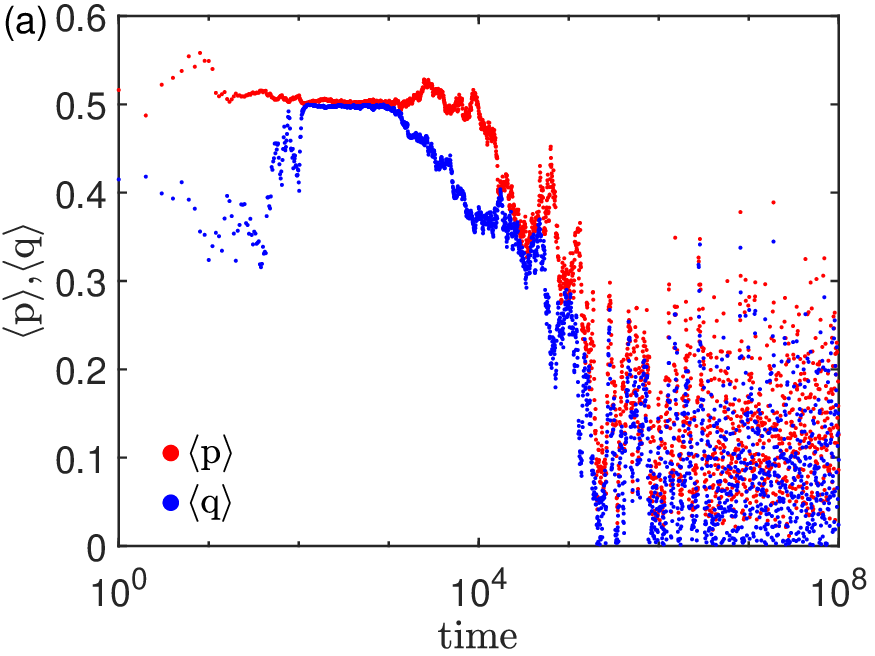}
\includegraphics[width=0.32\textwidth]{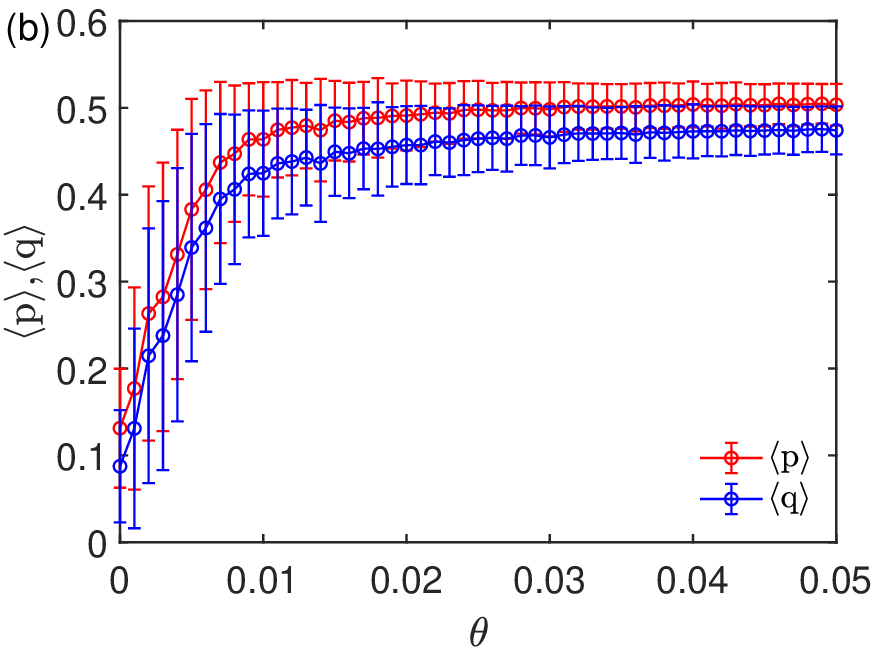}
\includegraphics[width=0.32\textwidth]{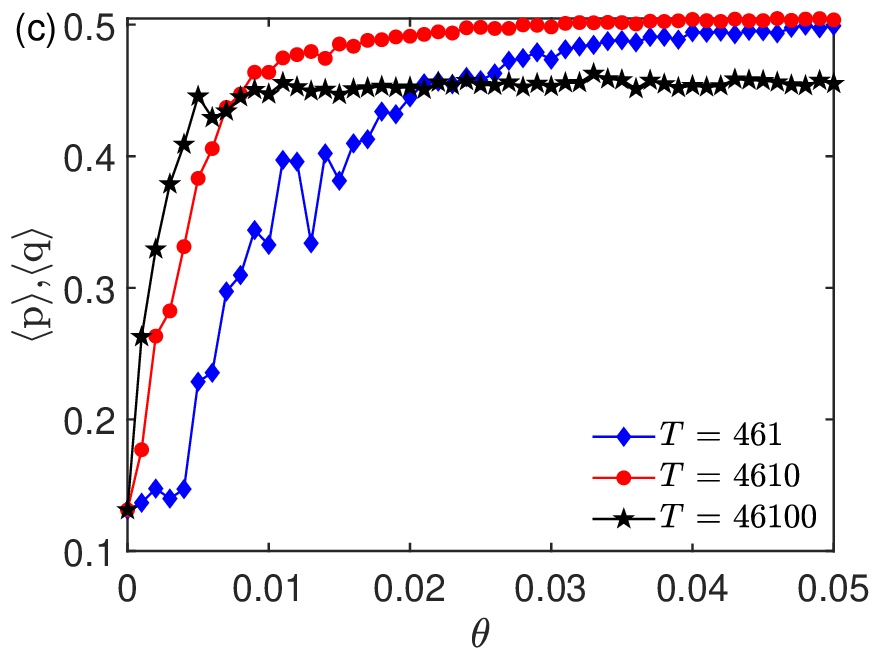}
\caption{(Color online) Periodic on-off pinning control of the largest hub in BA networks.
(a) The time evolution of $\langle p\rangle$ and $\langle q\rangle$ by pinning control till $10^3$ steps, afterwards the fairness evolves without pinning.
(b) The average offer $\langle p\rangle$ and acceptance $\langle q\rangle$ versus the parameter $\theta$, the period $T=T_c=4610$, the error bar represents their standard deviations.
(c) On-off pinning control with three different control periods $T$.
Parameters: $N=1000$, $\langle k\rangle=4$.
}
\label{fig:Appendix}
\end{figure*}
The two population structures studied in Sec. \ref{sec:results} are essentially homogeneous. In the real world, however, the underlying topologies are often heterogeneous, where the connectivity for different individuals could be very different. In a heterogeneously structured population, apart from the question of how many individuals are needed to be pinned, the selection of pinning nodes for the optimal control is also concerned.

Fig.~\ref{fig:BAtime}(a) provides an example that when $3\%$ nodes are randomly chosen to be pinned, the population evolves into a high fairness state on average, though intermittent failures are also seen. Fig.~\ref{fig:BAtime}(b) shows that when the pinning sites are chosen randomly, the level of fairness transition now becomes smooth, not sudden jump in the form of the first-order phase transition anymore as seen in the homogeneous cases. It shows that as the number of the pinning individuals increases till $3\%$ percentage of the population, the average level of fairness gradually approaches $S_h$. A larger pinning percentage further stabilizes the fairness level. 	

In fact, when only one hub with the largest degree is pinned, the fairness is still rapidly pinned to $S_h$ and is even more stable, see Fig.~\ref{fig:BAtime}(c). In numerical experiments, we find that even if the population size is increased up to $N=10^4$, pinning the largest hub is still sufficient to drive the whole population to the high fairness state.

To systematically investigate the impact of degree heterogeneity, we further give the minimal number of nodes $m$ versus the degree of the pinning nodes $k_i$. Fig.~\ref{fig:BAtime}(d) shows that quite some hub nodes alone rather than the largest hub can pin the level of fairness to approach $S_h$. Interesting, there is a power law for those small or intermediately large degrees $m\sim k_{i}^{-\gamma}$, with $\gamma\approx 1.4$. This means that to promote the fairness of heterogeneously structured populations, we can pin either very few hubs or relatively more periphery nodes according to the power law revealed. Approximately, the effect of pinning a node with degree $k_i$ is equivalent to pinning $10^{1.4}\approx25$ nodes with the degree of one order smaller $k_i/10$.

To understand the impact of the degree heterogeneity, it's helpful to derive the condition under which the fair strategy $S_h$ can spread for a given number of randomly pinning individual $N_p$. Let's consider a pair of interacting players $i$ and $j$, respectively with the strategy $S_h$ and $S_l$. Their expected payoffs are $\Pi_i=(1-C_h)+\frac{N_p}{N}(k_i-1)+(1-C_h)(1-\frac{N_p}{N})(k_i-1)$ and $\Pi_j=C_h+C_h\frac{N_p}{N}(k_j-1)+(1-\frac{N_p}{N})(k_j-1)$, respectively. The successful spread of $S_h$ is statistically conditioned by $\Pi_i>\Pi_j$, which leads to

\begin{equation}
\frac{k_i-2k_j+1}{2}+(\frac{k_i}{2}+\frac{k_j}{2}-1)\frac{N_p}{N}\geq0,
\label{eq:sf}
\end{equation}
where $C_h=0.5$, the same as in the above. The larger of the left side of this inequality, the more likely for the fair strategy $S_h$ to spread. While the first term could be either positive or negative, depending on the $k_{i,j}$, however, by plotting its value versus degree $k_i$, it can be found that the probability of the first term being positive increases with the increase of $k_{i}$. The second term is always positive and is larger if node $i$ is a hub. This explains why pinning hub node is more efficient than the periphery nodes as shown in Fig.~\ref{fig:BAtime}(d).

\subsection{On-off pinning scheme}
\begin{figure}
\centering
\includegraphics[width=0.45\linewidth]{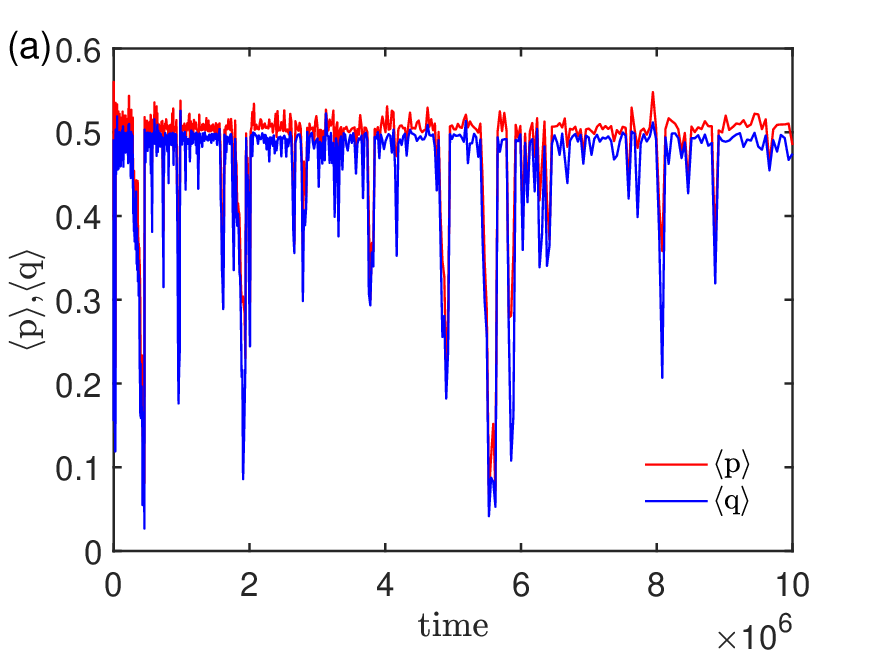}
\includegraphics[width=0.45\linewidth]{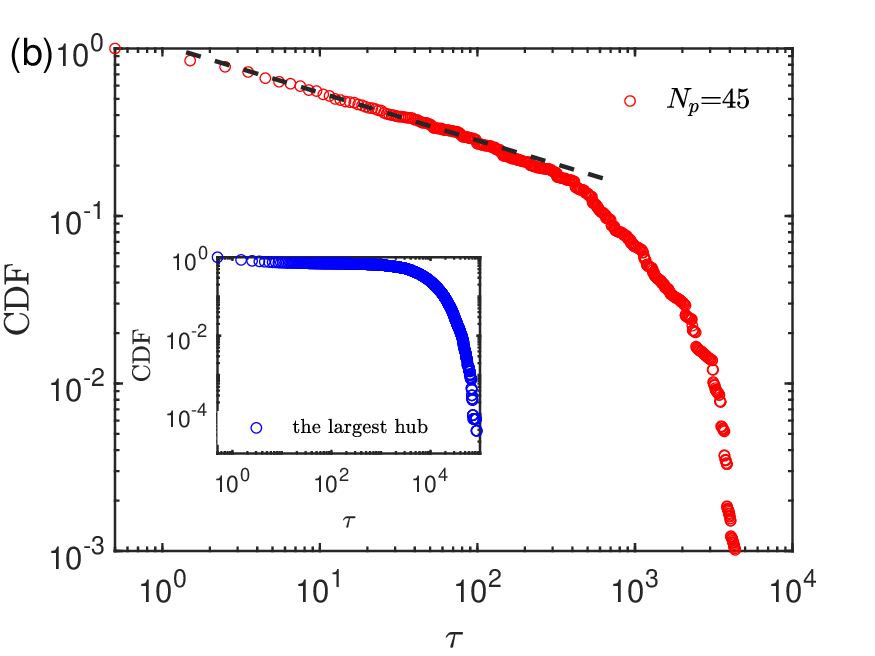}
\caption{(Color online) Intermittent behaviors in BA scale-free networks.
(a) Typical time series of intermittent failures when $45$ nodes with degree $2$ are pinned.
(b) The corresponding cumulative distribution curve (CDF) of the interval time $\tau$ of intermittent events, the inset is for the scenario of pinning the largest hub. The black dashed line is a fitting line with an exponent of $-0.35$.
Parameters: $N=1000$, $\langle k\rangle=4$.
}
\label{fig:ERSF}
\end{figure}

In order to further reduce the control cost, we find a periodic on-off pinning control~\cite{Chen2009Synchronization} for the largest hub is sufficient to sustain the fair state $S_{h}$ and we implement it as follows.

First of all, we compute the characteristic time $T_{c}$ of the system as the control period. Specifically, we adopt the idea of the finite-time Lyapunov exponent in the nonlinear dynamics~\cite{Strogatz2015Nonlinear}
by analogy as follows
\begin{equation}
\lambda=\frac{1}{\Delta t}\ln (\frac{|\Delta p'|}{|\Delta p|}),
\label{eq:pi1}
\end{equation}
$\Delta p=0.5-\langle p(t)\rangle$ denotes the distance to the high fairness state $S_h$ at some moment $t$ when the system within pinning control reaches the steady-state. When the pinning control is withdrawn, the fairness is expected to decrease. $\Delta p'$ is the new distance to $S_h$ after $\Delta t$, $\Delta p'=0.5-\langle p(t+\Delta t)\rangle$, see Fig.~\ref{fig:Appendix}(a). In our computation, $t=1000$, $\Delta t=10000$, the finite-time Lyapunov exponent $\lambda$ is averaged 100 realizations. From the nonlinear dynamics point of view, the pinning control acts as a perturbation, the value of $\lambda$ reflects the evolution rate back to the low fairness state, captures the time-scale of the dynamics. Once the Lyapunov exponent $\lambda$ is obtained, the characteristic time $T_{c}=\frac{1}{\lambda}$, and $T_c\approx 4610$ by our computation.

Denote $\theta$ as the proportion of the period within which the pinning control is periodically turned on. In Fig.~\ref{fig:Appendix}(b), we can see that the population has steadily risen to the high fairness when $\theta\gtrsim0.02$. It means that the high fairness of the whole population can be achieved with only about 1/50 control time windows. The standard deviation, however, shows that a larger $\theta$ is required if smaller fluctuations are desired. This indicates that due to the structural heterogeneity, discontinuous pinning may be a good option to reduce the control cost. Finally, we also compare the results with different period $T$, showing that the characteristic time $T_c$ we adopted is a reasonable choice, a longer or shorter period yields worse results, see Fig.~\ref{fig:Appendix}(c).

\subsection{The intermittent behaviors}
In the end, we study the intermittent behaviors as we have already seen in Fig.~\ref{fig:BAtime}(a). A better example is shown in Fig.~\ref{fig:ERSF}(a), where $45$ nodes of degree $2$ are pinned. The fairness level drops from time to time even when the high fairness state is reached, and these intermittent failures are present for arbitrarily long time evolution. We define each $\tau$ as the time interval during which $\langle p\rangle<0.5$ and do a statistics of this time duration $\tau$ for these intermittent failures. The CDF indicates that there is a power law with an exponent $-0.35$ for about two orders in $\tau$, with a heavy tail at the  end. This implies that the system is near some sort of critical state~\cite{Helmut1989Intermittency,O.B2005A,Hammer1994Experimental,Zhuo2021Characteristics,Detlef1993ntermittency,Huepe2004Intermittency}. The observation disappears when less or more nodes are pinned, or the hub nodes are pinned, as shown in the inset of Fig.~\ref{fig:ERSF}(b), as well as Fig.~\ref{fig:BAtime}(c).

\section{Conclusion and Discussion}
\label{sec:discussion}

In summary, we apply the idea of pinning control to the UG model and show that this is an efficient approach to promoting social fairness. For homogeneously structured populations, we reveal that pinning a very small fraction of individuals to be fair players can drive the overall fairness of the population. Though the converging time to the high fairness state shows quite different dependence on the pinning number for $1d$ lattice and ER networks.
For heterogeneously structured populations, this leverage effect is even more pronounced that a single hub is sufficient to do so. A power law is revealed for the minimal number required to achieve the fair state for pinning nodes with different degrees. We also provide some analysis of why pinning hub works better. But, if the pinning control is marginally strong, intermittent failures can be seen.

Our analysis shows that once the pinning state is reached, the payoff difference between the pinned individuals and the unpinned is small, meaning no more control cost is needed afterwards. But if failed, the payoffs for the pinned are indeed lower than the rest, which might need to be compensated. For the heterogeneously structured population, one convenient option is to pin only one hub. To further reduce the control cost, the control can even be imposed discontinuously, e.g., a periodic on-off control procedure can achieve the goal.

Within the paradigm of \emph{Homo economicus}, the behaviors sticky to the high fairness might be considered irrational, which can only be explained by supported by some external support for compensation.
In our daily lives, however, there are indeed some ``good Samaritans" that their decision-making is not only rationality-driven but also morality- or emotion-driven. In these scenarios, fair behaviors are spontaneous that need no compensation. Anyway, in those societies where the fairness level is unsatisfactory, our study indicates that the pinning control might be a feasible option to improve.

\ack
We are supported by the National Natural Science Foundation of China under Grants Nos. 12075144 and 12165014.

\section*{Appendix}
Without loss of generality, we assume that the strategies of all individuals of well-mixed population $N$ settle down around a solution $s=(p, q)$, now we show that if $p\neq q$, the strategy must be evolutionally unstable.

If $p > q$, an individual with the mutant strategy of $s_1=(p_1,q$) with $q<p_1< p$ can invade the population with the strategy $s$. Because the two payoffs are respectively $\pi_{s}=(N-2)+(1-p+p_1)=N-1+p_1-p$ and $\pi_{s_1}=(N-1)(1-p_1+p)$, where we have $\pi_{s_1}-\pi_{s}=N(p-p_1)>0$. This means that the value of $p$ for those individual holding $s$ will decrease until it approaches $q$.

If $p<q$, an individual with the mutant strategy of $s_2=(p_2,q)$ with $q<p_2<1$ can invade the population with the strategy $s$. Because the payoffs for the individuals holding $s_2$ and $s$ are respectively $\pi_{s_2}=(N-1)(1-p_2)$ and $\pi_{s}=p_2$, where $\pi_{s_2}>\pi_{s}$, if this happens, the $p$ value of the individual with the strategy $s$ in the population will go up till the value of $q$. Similarly, an individual with the mutant strategy of $s_3=(p, q_3)$ with $0<q_3<p$ can also invade this population, because the payoffs are respectively $\pi_{s_3}=(N-1)p$ and $\pi_{s}=1-p$, where we also have $\pi_{s_3}>\pi_{s}$. As time goes by, the value of $q$ for those individual holding $s$ will decrease till the value of $p$.

Therefore, the stable strategy in the end $s=(p, q)$ must satisfy $p=q$. But due to the small mutations $(\delta p,\delta q)\in[-\epsilon,\epsilon]$ in the model, $p$ will be slightly larger than $q$ to guarantee the division agreement by evolution.

\section*{References}

\bibliographystyle{iopart-num}
\bibliography{fairness}   
\end{document}